\documentclass{AIAA}

\begin{document}

\title{iMapD: intrinsic Map Dynamics exploration for uncharted effective free energy landscapes}

\author{Eliodoro Chiavazzo\footnote{eliodoro.chiavazzo@polito.it}}
\affiliation{Energy Department, Politecnico di Torino, Torino 10129, Italy}
\author{Ronald R. Coifman\footnote{coifman@fmah.com}}
\affiliation{Department of Mathematics, Program in Applied Mathematics, Yale University, New Haven CT 06510, USA}
\author{Roberto Covino\footnote{roberto.covino@biophys.mpg.de}}
\affiliation{Max Planck Institute of Biophysics, 60438 Frankfurt am Main, Germany}
\author{C. William Gear\footnote{wgear@princeton.edu}}
\affiliation{Department of Chemical and Biological Engineering, Princeton University, Princeton, NJ 08544, USA}
\author{Anastasia S. Georgiou\footnote{tasia.georgiou12@gmail.com}}
\affiliation{Department of Chemical and Biological Engineering, Princeton University, Princeton, NJ 08544, USA}
\author{Gerhard Hummer\footnote{gehummer@biophys.mpg.de}}
\affiliation{Max Planck Institute of Biophysics, 60438 Frankfurt am Main, Germany}
\affiliation{Institute of Biophysics, Goethe University, 60438 Frankfurt am Main, Germany}
\author{Ioannis G. Kevrekidis\footnote{yannis@princeton.edu}}
\affiliation{Department of Chemical and Biological Engineering, Princeton University, Princeton, NJ 08544, USA}
\affiliation{PACM, Princeton University, Princeton, NJ 08544, USA; also IAS-TUM, Garching, Germany}

%\keywords{Free energy surface $|$ Model reduction $|$ Machine learning $|$ Protein folding}

\begin{abstract}
We describe and implement iMapD, a computer-assisted approach for accelerating the exploration of uncharted effective Free Energy Surfaces (FES), and more generally for the extraction of coarse-grained, macroscopic information from atomistic or stochastic (here Molecular Dynamics, MD) simulations.
The approach functionally links the MD simulator with nonlinear manifold learning techniques.
The added value comes from biasing the simulator towards new, unexplored phase space regions {\em by exploiting the smoothness} of the (gradually, as the exploration progresses) revealed intrinsic low-dimensional geometry of the FES.
\end{abstract}

%\keywords{Free energy surface $|$ Model reduction $|$ Machine learning $|$ Protein folding}

\maketitle

\section{Introduction}
A crucial bottleneck in extracting systems-level information from direct statistical mechanical simulations in general, and from Molecular Dynamics (MD) in particular, is that the simulations sample phase space "at their own pace", dictated by the shape and barriers of the effective free energy surface. 
Long simulation times are thus "wasted" revisiting already explored locations in conformation space.
Over the last twenty years there has been a tremendous amount of effort invested, and many truly creative solutions proposed, for biasing the simulations so as to circumvent this. 
Techniques that have now become a standard part of the simulator's toolkit, like umbrella sampling or SHAKE, biasing techniques like importance sampling, milestoning, path sampling or metadynamics, techniques like the nudged elastic band/string method, and -more recently- techniques based on machine learning, like reconnaissance metadynamics, have been ingeniously formulated and proposed to help alleviate this problem (see e.g. \cite{Abrams2013,Spiwok2015,E2010,Jung2012,milestoning} and references therein). A recent review on dimensional reduction and enhanced sampling in atomistic simulations can be found in \cite{Rohrdanz2013}.

A crucial assumption that underpins many of these methods (as well as our own work over the years, and what will be implemented here) is that the dynamics are, effectively, low-dimensional:
there exists a "good set of a few collective variables" (also called collective coordinates, or reduction coordinates) in which one can write an effective Langevin (or effective Fokker-Planck) equation; it is the potential of this effective Langevin that we are trying to identify and exploit.
While one generally expects this effective Langevin to be a higher order, generalized one, and thus have memory terms \cite{Baker2016}, we will show here how we can, in effect, construct "short memory"
approximations with the help of collective variables detected and
updated "on the fly" using manifold learning.
If we (a) knew the right coarse-grained observables (the right collective variables); and if importantly, (b) we had an "easy way" to create molecular conformations consistent with given values of these coordinates (a procedure that we call here "lifting"), then creating tabulated (or interpolated) effective free energy surfaces with a black box atomistic simulator and umbrella sampling would be "easy".
By observing the dynamics of the MD in these few collective coordinates, we can then straightforwardly estimate the local gradient of the effective potential and the local diffusivity in the effective Langevin description. 
Actually, "easily" does not do justice to the problem - estimating effective Langevin terms locally from simulations is a highly nontrivial estimation problem in the theory of SDEs, and many a career in financial mathematics are made from studying it carefully. 
Here, we will conveniently assume that we have at our disposal "the current best" local stochastic estimation techniques available, so that we can go from observations of the unbiased dynamics (or the umbrella-sampled dynamics) to local effective SDE term coefficients.

Given an approximate effective free energy surface (in its "few" collective coordinates), we can then go ahead to perform tasks like reaction rate estimation with the explicit surrogate function, or its tabulated form. 
Mathematical and computational tools for performing such tasks on explicit (or tabulated) functions of a few variables exist in the "standard" mathematical literature (e.g. in optimization) and will also be assumed known and "off the shelf" available.

Even though finding these invaluable good collective coordinates is difficult, it is at least reassuring to know that "the useful collective coordinate set" is not unique, but rather conveniently degenerate: any set of two basis vectors on a plane suffices to span the plane.
If our -say for simplicity- two-dimensional free energy surface is not a flat plane, but a curved manifold, any set of two basis vectors on any plane that is one-to-one with our FES would suffice to parametrize it (provide a set of coordinates on it, so that we can navigate it).
%%%%%%%%%%%%%
%
%
\begin{figure*}%[tbhp]
\centering
\includegraphics[width=1.0\linewidth]{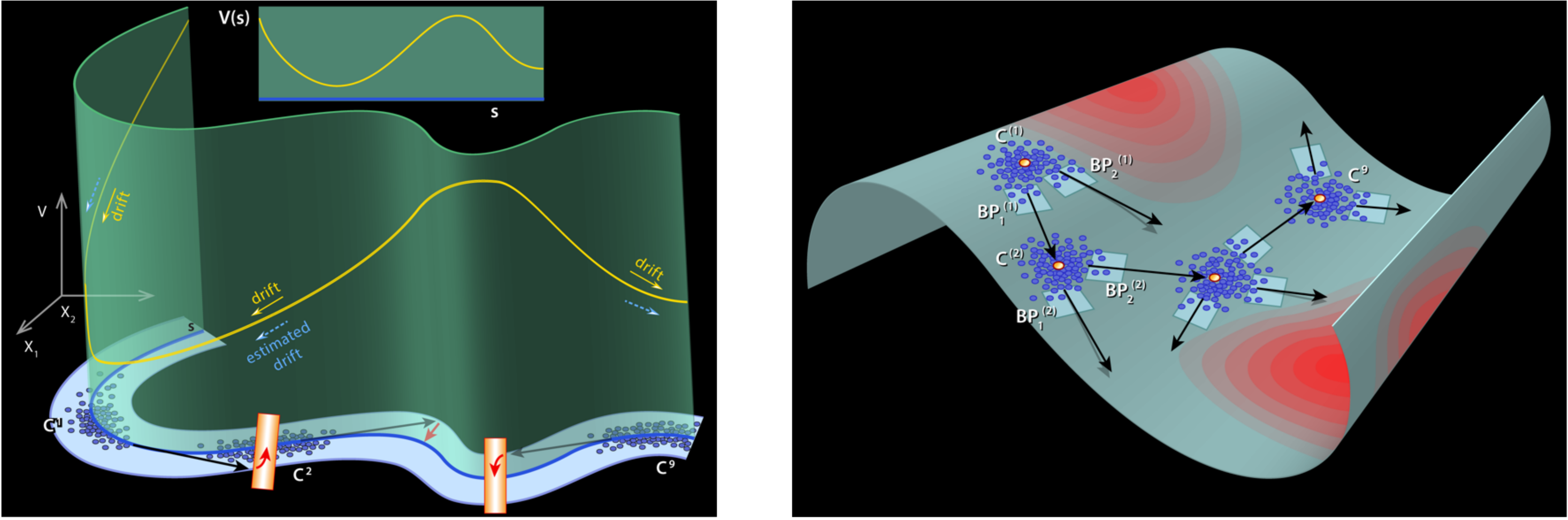}
\caption{Pictorial illustration of the iMapD exploration procedure with one-dimensional (left panel) and two-dimensional (right panel) effective FES. In the top left inset, a good collective coordinate (s) is already available - the collective coordinates in the main panels are not {\em a priori} known. See full description in the text.}
\label{draft01}
\end{figure*}

%%%%%%%%%%%%%%%%%%%%%%%%%  22/08/2016
Discovering good coordinates for describing a function based on data lies at the crux of modern computer/data science research.
This is precisely our task here, too: find (explore, reveal) an effective FES, and parametrize it (construct a map of it) in terms of useful collective coordinates.
We first discuss the simple,  one-dimensional case. 
If a good, physically meaningful reaction coordinate $s$ is known {\em a priori} (see the inset in left panel of Figure \ref{draft01}) then a procedure for extracting a good approximate FES from computational data is obvious: a (possibly regular) grid in the one-dimensional phase space is constructed, and umbrella sampling is performed to estimate the potential of mean force. 
Alternatively, several parallel, appropriately initialized short runs can be used to estimate the local effective Langevin drift and diffusivity. 
Either way, an approximate FES (with controlled approximation error) can be interpolated.
%

%%%%%%%%  22/08/2016
Yet good collective coordinates are not globally known in advance, and must be generated as the computation progresses. 
Consider as an illustration, a one-dimensional manifold (blue thick line in Figure \ref{draft01}) with the corresponding 1d effective FES (in yellow).
Say that after some initial simulation time the solution trajectory becomes trapped in one of the energy wells (bottom left in the figure).
Then data mining can be applied to an ensemble of locally sampled configurations to (a) establish that the relevant manifold is one dimensional; to (b)
learn its local parametrization  (e.g. in terms of the first meaningful diffusion map coordinate); and thus to (c) detect the boundary points of the manifold portion so far explored ("fathomed").
These boundary points can now be smoothly extended outward - not in time, but in the geometry of the manifold, parametrized locally by the first diffusion coordinate \cite{Coifmanpaper1}, or, even simpler, by the first local Principal Component close to each boundary.
This extension takes us beyond the conformation space already explored, and may well be {\em against} the local FES gradient, leading, as we will discuss, to computational savings.
The extension cannot but be an approximate one - a "predictor", a Taylor series approximation of the manifold locally in the ambient space (black arrow on the left in the left panel of Figure \ref{draft01}).
A "corrector" step must follow: a short equilibration (possibly by PLUMED-assisted umbrella sampling \cite{plumed01} as schematically illustrated by the red arrow following the black one in the figure; or possibly by a short unbiased simulation) gives us new, unexplored conformations on the manifold beyond what we had already fathomed.
New brief simulations are run, new points on the manifold are collected (second blue point cloud in the figure), added to the data base, and then fed to data mining so as to parametrize the augmented FES geometry.
The extension procedure repeats again and again: new unexplored conformations keep being added and the extended geometry of the effective FES is gradually revealed, leading to the discovery of new wells (like the one on the right in the left panel of Figure \ref{draft01}).

%%%%%%%%  22/08/2016
In higher dimensions, the basic approach remains the same, although its representation becomes more complicated.  
Consider a two-dimensional "undulating" free energy surface embedded in a high dimensional ambient space, as shown in the right panel of Figure \ref{draft01}. 
The "color map" on this carpet denotes the effective FES contours; if, as we "walk" on the carpet we estimate, on the fly, the local color gradient, we can use this information to help us direct our walking pattern, and thus make the extraction of the information we are after (like a reaction rate, or the discovery of a saddle point) less time consuming.

In "Indiana Jones and the Last Crusade" the hero walks on a glass mirror bridge that he cannot see - but in the end, he takes some sand, and throws it at his feet, so that the sand reveals the local shape of the bridge. 
This is precisely what we do in iMapD with our "free energy carpet". We start with simulations that have locally (and partially) sampled some location on the FES: "sand we have poured around our own feet" - our first small cloud, "C1" in right panel of Figure \ref{draft01}; but now that a little of the low-dimensional geometry of the carpet is revealed, we can "walk", take a big step (as big as we trust the smoothness of the carpet) to a new location, and pour some sand there, namely initialize molecular dynamics conformations consistent with the new location in collective coordinate space, and start one or more unbiased simulations there.
This is "the new sand" - our new little cloud, C2 in the right panel of Figure \ref{draft01}, to which we have stepped. 
One can easily intuit how the geometry is revealed from iterating this process: For a  $d$-dimensional (here $d=2$) reduced description, the initial "seed" simulation will form an effectively $d$-dimensional (here $2$-dimensional) cloud. 
We need to identify the $(d-1)-$dimensional (here, $1$-dimensional) boundary of this cloud (its "silver lining"). 
Then, at some point on this lining, -or possibly at an ensemble of points uniformly discretizing this $1$-d curve, each of them one by one, or all in parallel computationally- we "take a step away" from this cloud, smoothly extrapolating the coordinates "as far as we trust their smoothness". 
We could move, for example, along local geodesics -as we have explained in detail in previous work \cite{ThomasPaper}; clearly, we can only "trust" these geodesics only so far (since the carpet may "violently" curve, and our smooth extrapolation may not locally parametrize it any more).

By identifying the boundary, marching "outward" from a number $M$ of points on it, creating $M$ little $d$-dimensional clouds from simulations initialized at each extrapolation, and then "integrating" the new $M$ clouds in an atlas with the initial one, again and again, we will "fathom" the carpet. 
So, while the "local marching" from every boundary parametrization point may be done in local coordinates (e.g. local PCA) or in more global coordinates (e.g. diffusion maps geometric harmonics), all new data points at every iteration can be integrated in our global geometry either by reprocessing them all together, or by creating an efficient database structure that allows transitioning from local coordinates of one cloud to local coordinates of the cloud next to it (one chart in an atlas to the next). 
This is reminiscent of "simplicial continuation" in following the parametric solutions of  algebraic equations.
We have a predictor step (our extrapolation in local reduction coordinates) and then a corrector step that "brings us back down" to the FES - this corrector step might be discarding the fast initial transient of an unbiased simulation, or running umbrella sampling constrained on the extrapolated coarse coordinates.

It is clear that we only need to move "outward" if we are to explore new areas - (that can be ensured through good bookkeeping) - and while this is a nontrivial process to generalize, bookkeeping is easy in one dimension, easy (but nontrivial) in two dimensions, doable in three, difficult in four etc. - so practically, we expect the process to be easy to program for relatively low-dimensional (1,2,3... maybe 4 dimensional) free energy surfaces - and it will require nontrivial  computational geometry and programming.
This is precisely the same bookkeeping necessary in multi-parameter simplicial continuation for the tracking of solutions of algebraic equations \cite{COCO} (see also the available software package for continuation and bifurcation problems AUTO \cite{AUTO}).

\section{Results}
We discuss now the implementation of the proposed iMapD molecular simulation sampling approach. 

Conforming to the literature (but also for the sake of clarity) our first benchmark illustration is the time-honored alanine dipeptide  \cite{ThomasPaper,CeciliaPaper,GerhardYannis}, here in implicit solvent with the Amber03 force field \cite{amber03}.

In our second application, we apply the iMapD algorithm to the transmembrane protein Mga2, which plays a key role in the regulation of lipid saturation levels in the yeast endoplasmic reticulum (ER).  Recent simulations and experiments identified a unique rotation-based  sensing mechanism to probe the membrane characteristics \cite{Covino2016}. In response to changes in lipid saturation, the 30-amino acid transmembrane helices (TMH) anchoring Mga2 into the ER were found to rotate relative to each other in an Mga2 dimer, driven in part by packing effects acting on bulky protruding  tryptophans. Just probing the rotational dynamics and charting the underlying free energy landscape required  millisecond-long MD simulations feasible only with a coarse-grained (CG) description \cite{Marrink2007,Monticelli2008}. However, even on this long time scale, only the TMH contact could be sampled, with TMH dissociation expected to occur on time scales orders of magnitude longer. Therefore, even in more than 3-ms simulations of a simplified CG description, the relevant configuration space of the dimer could not be sampled exhaustively.

Here we show that with iMapD not only the competing Mga2 bound states, but also the unbinding pathways can be discovered, simply by strategic initialization of otherwise fully unbiased MD trajectories. This reduces computational costs at least 1000-fold. 
\subsection*{Benchmark 1} 
\begin{figure*}
\centering
\includegraphics[width=1.0\linewidth]{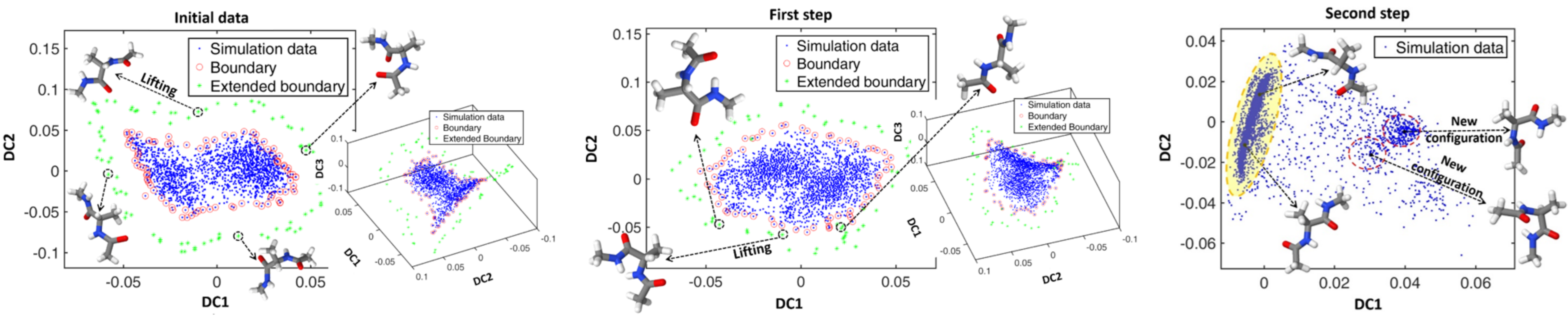}
\caption{\textbf{Left panel}: A long initial trajectory trapped in two nearby metastable wells is shown in the corresponding two-dimensional DMAP projection (a three-dimension DMAP space embedding is also reported in the inset on the right-hand side). Boundary points are identified (red circles) and extended outward (green stars). Local PCA suffices to lift to ambient space (see below). \textbf{Middle panel}: Short simulation runs are performed from previously extended boundary points. New configurations are generated and displayed (blue dots) in a two-dimensional (and in a three-dimensional) DMAP reduced space. \textbf{Right panel}: After two steps, two new potential wells are reached by some of the simulation frames. The ``starting" portion of the FES geometry accessed by the initial simulations is marked in yellow - the rest has been revealed through exploration.}\label{AlaMulti}
\end{figure*}

It has been long argued that alanine dipeptide admits a two-dimensional reduced description in terms of two physically meaningful coordinates, namely the dihedral angles $\phi$ and $\psi$ (see details, e.g., in \cite{GerhardYannis}).
Our approach does not require such {\em a priori} knowledge (neither of the dimensionality nor 
of some  physical meaning of the collective coordinates) while exploring the FES.
Three successive stages of our exploration protocol are reported in Figure \ref{AlaMulti}; the protocol is initialized from a transient simulation segment (an ensemble of configurations) visibly trapped within some initial potential well; this is what we call the {\em initial simulation data}.
Each stage of iMapD is composed of the following sub-steps:
\begin{itemize}
 \item {\bf Data Mining}. A manifold learning technique (here, Diffusion Maps - DMAPs -  \cite{Coifmanpaper1}) is used to discover a low-dimensional embedding for the data collected so far. This discovery includes the selection of the appropriate dimension ($d$) of the manifold, and its parametrization, here in terms of $d$ leading diffusion map coordinates (DC1,...,DCd).
\item {\bf Boundary Detection}.  Using algorithms from the literature (e.g. here alpha-shapes \cite{ashapes1,ashapes2}, readily implemented in the Matlab package, or more generally, ``wrapping" algorithms \cite{Edelsbrunner2003})
we detect the $d-1$ dimensional boundary of the region explored (``fathomed'') by the available simulation data.
\item {\bf Outward Extension}. At each boundary point, we take an outward step (here, approximately normal to the boundary, in the tangent space of our low-dimensional manifold). The option implemented in this study involves Local Principal Components (LPC) in the ambient space. For each boundary point, (a) a fixed number of nearest neighbors is detected in ambient space; (b) local PCA is performed on this set of neighbors with the local reduced dimension $d_{loc}$ selected by a threshold for the maximum variance (details below); (c) the center of mass of this local neighborhood is computed in the $d_{loc}$-dimensional PCA space; and then (d)
``outward extension'' of the manifold at the original boundary point is performed (in the PCA low-dimensional space) along the line segment passing through the local neighborhood center of mass and the boundary point itself. 
Alternative extension techniques (like geometric harmonics or
Laplacian pyramids \cite{LafonTh,ProcessesPaper} can also be used
for this purpose.
\item {\bf Lifting} is then performed from the extended LPC coordinates to novel, unexplored  molecular configurations lying on/close to the extended manifold. Going from LPC directly to ambient space was satisfactory in this simple illustration. In general, however, equilibrated conformations consistent with ({\em respecting}) the extended LPC coordinate values 
may be needed, and
can be obtained, for example, through short, constrained, umbrella-sampling runs \cite{GerhardYannis}, e.g. through PLUMED \cite{plumed01} or Colvars \cite{Colvars}.
\item{\bf New sampling/data base updating on the extended manifold}. Short simulation bursts are carried out from these new ``extended" initial conditions (possibly several replicas from each, initialized with different Maxwell-Boltzmann velocities and/or different thermostat seeds).The new data are appended to the growing fathomed configuration data base.
\end{itemize}
The procedure then repeats till new metastable configurations are detected.

In Figure \ref{AlaMulti} (left panel) an initial transient is visibly trapped in two nearby metastable configuration wells. 
Data mining (DMAPs) clearly suggests a two-dimensional FES ; the inset shows this two-dimensional
manifold embedded in the the space of the first three DMAP coordinates.  
The boundary points of the ``fathomed" portion of the manifold are identified (red circles) and extended outward (green stars).
Lifting via LPC is quite satisfactory here, and new sampling on the extended manifold is performed through
simple unbiased short runs initialized at the lifted configurations.
The resulting new configurations are appended to the growing simulation database, and a new round of data mining, boundary detection and outward extension is shown in Figure \ref{AlaMulti} (middle panel), both in 2d projection and
in 3d embedding.
This is repeated one more time, leading to Figure \ref{AlaMulti} (right panel), where two new (folded) metastable configurations have been discovered.
What is important is that the manifold parametrization shown in the right panel of Figure \ref{AlaMulti} {\em was not known} at the beginning. 
Only the small portion of the manifold (marked by the yellow ellipse) was initially available.
The geometry of the (growing) manifold beyond that initial ellipse and its (adaptively also growing) parametrization have been gradually revealed as part of our exploration protocol.

\begin{figure}
\centering
\includegraphics[width=0.5\linewidth]{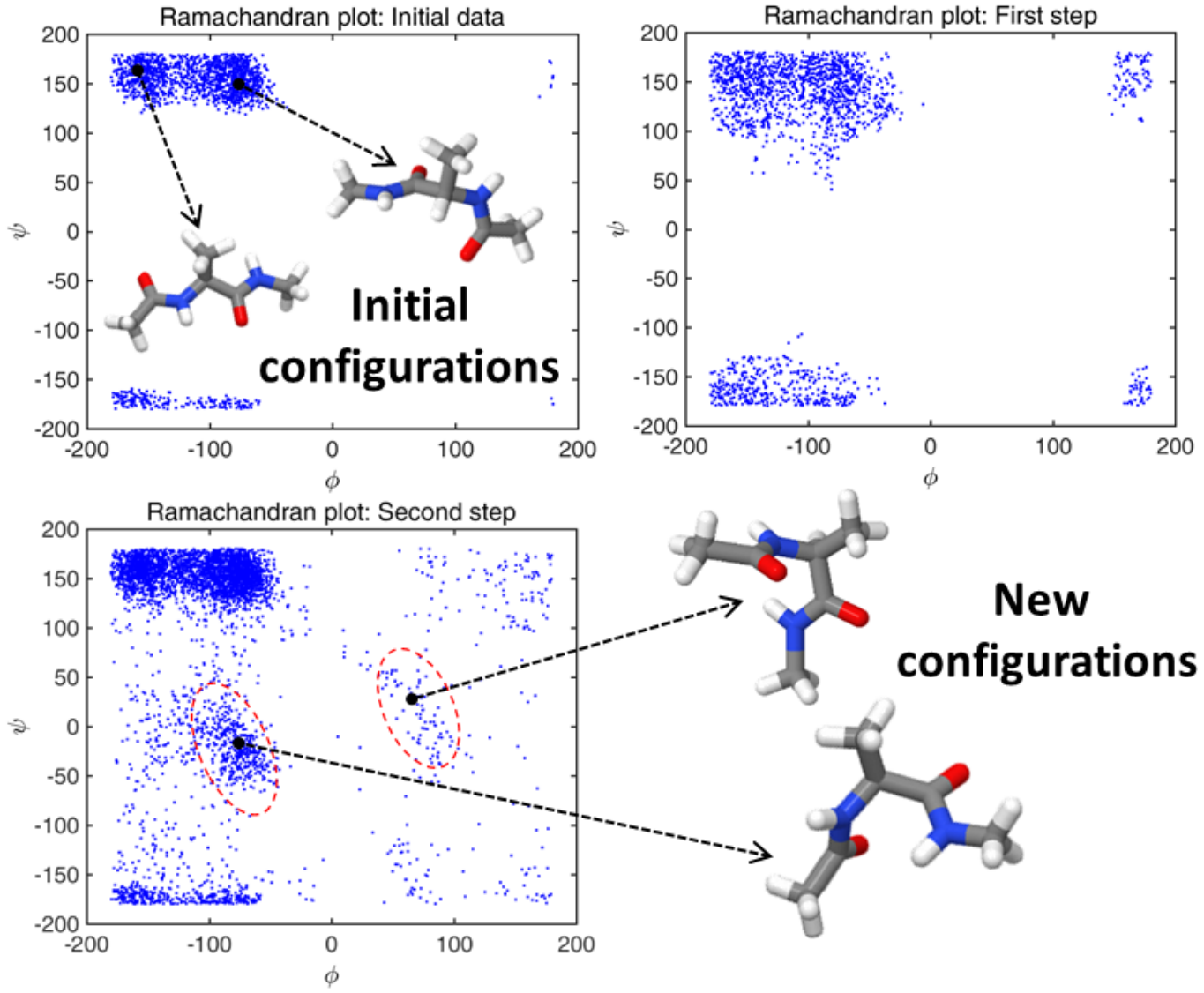}	
\caption{The discovery process of the above Figure is redisplayed here as a Ramachandran plot. Two steps are sufficient to reveal two initially unknown metastable configurations of the molecule. At each step, before outward extension of the boundary, we also performed global PCA filtering of the data noise where $98\%$ of the variance was retained (see Materials and Methods below).}\label{Discovery}
\end{figure}

%----------------LATER
For the sake of completeness, the results of the above exploration process are also reported in the popular Ramachandran plot in Fig. \ref{Discovery} in terms of the two physical coarse variables $\phi$ and $\psi$. 
The dipeptide, initially trapped in the basins on the top-left, is gradually forced towards new configurations that would not have naturally been visited in such a short simulation time period.
In this relatively simple example, the low-dimensional FES ``slow manifold'' identified on the fly happens to also be the graph of a function above the two Ramachandran plot coordinates (in other words, the determinant of 
the Jacobian of the transformation form the ``physical" coarse coordinates $\phi$-$\psi$ to the diffusion ones DC1-DC2 keeps the same sign, and is neither too big nor too small on the data: it stays bounded away from zero and from infinity, so that the transformation from physically meaningful to data-based collective coordinates is bi-Lipschitz \cite{Ronen}).
This means that the effectively 2d FES can be described {\em equally well} in terms of $\phi$-$\psi$ or in terms of our (evolving) DC1-DC2. %$\xi_1$-$\xi_2$.
If, however, this effectively 2d manifold ``folded" over the Ramachandran plot variables, our data-mining would still be able to correctly parametrize, and extend, the FES.

Before we elaborate on the steps, a few words about efficiency.
In this example, the total computational time associated with all performed simulation bursts was estimated at $\approx 50$ ps.
It is known that for this system $\approx 150$ ns direct simulation are, on average, needed to observe the transition from the initial, lower free-energy configurations to the discovered, higher free-energy ones (see \cite{CeciliaPaper} where the same system
% GH: above we write "in implicit solvent"
%in vacuum with Amber03 force field
was simulated). 
This yields an apparent computational speed-up of three orders of magnitude for this rudimentary (far from optimal) implementation, in line with what was observed in \cite{CeciliaPaper}, where the re-initialization did not involve extrapolation, but rather occurred at the ``farthest reached'' point (in DMAP space); since the computation there was one-dimensional, boundary detection was straightforward.
This technique, like reconnaissance metadynamics, also builds the exploration geometry, and actually does it "seamlessly", without having to "jump and reinitialize" consistent molecular configurations. 
Yet it is precisely this ``jumping and reinitializing'' that we feel is the most powerful element of our approach: we do not have to wait to ``fill in the wells'' (as in metadynamics), nor do we need to sample the part of the geometry that we trust is smooth enough. 
We can take a step ``as big as we trust'' in the geometry, and then sample there and in this way save a remarkable amount of computational time (see the Appendix). 
These two steps are determined by the length $\Delta t$ of the unbiased sampling and the length $c$ of the extension. Before discussing optimal choices of these two crucial parameters, some conceptual geometrical considerations are in order.

For a simple one-dimensional SDE in terms of a known variable $x$, reinitializations can be carried out with no effort and the extension parameter $c$ chosen as large as one likes (the support of the effective FES is not a curved one-dimensional manifold in a high dimensional space).
In complex atomistic simulations, however, challenges to the practical implementation of the above procedure arise because
\begin{itemize}
\item the low-dimensional {\em support} (manifold) of the effective free-energy surface is typically curved and embedded in a high dimensional phase space;
\item coarse grained coordinates parametrizing this manifold are {\em a priori} unknown and need to be systematically discovered and ``harmonized" with their incarnations at the previous step;
\item re-initialization of the fine-scale simulator requires a {\em lifting} operator from the low-dimensional space up to ambient physical space (see also \cite{ProcessesPaper}).
\end{itemize}
Those are precisely some of the aspects addressed in this work.
Our Alanine example only provides a proof-of-concept illustration since the code we set up was far from optimal. 
We did not optimize the extension parameter $c$ nor the unbiased sampling time $\Delta t$, which was instead kept constant; and in this first attempt, at each step we extended {\em all} detected boundary points.
Future optimized implementations will include a smarter parametrization/selection of the boundary points to extend, as well as a smarter selection of the unbiased sampling interval based on local estimates of the free energy gradient (this selection follows the same principles discussed in
detail and demonstrated in \cite{ThomasPaper}, where, however, the collective coordinates were already known).
One might, for example, not extend points at which the effective FES rises steeper (its local gradient norm is larger) than a preset threshold.
Finally, in estimating computational speedup we should also include the cost of necessary intermediate steps, such as DMAPs, local PCA and lifting.
\begin{figure*}%[tbhp]
\centering
\includegraphics[width=1.0\linewidth]{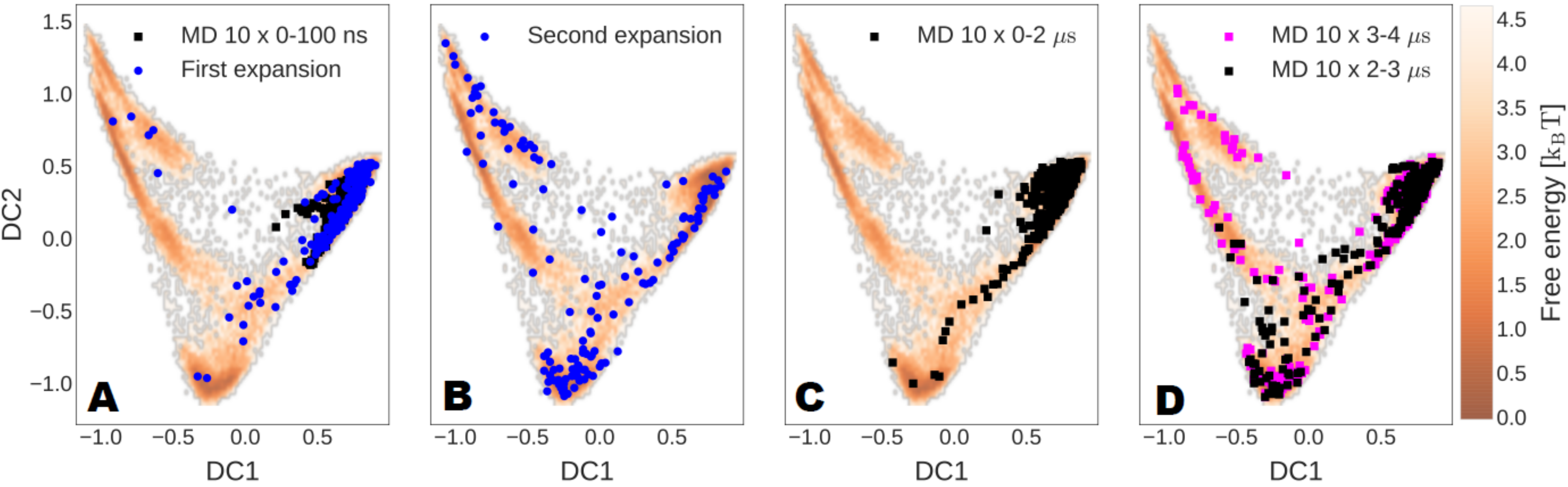}
\caption{Enhanced exploration of Mga2 dimer configurations represented on the free energy surface as a function of the first two global diffusion map coordinates,  DC. (\textbf{A}) Configurations sampled from ten 100 ns long unbiased simulations initiated from a single configuration (black squares). Final configurations of 100 ns long unbiased simulations initialized  from the first set of 16 {\em newly projected} structures (blue circles). (\textbf{B}) Final configurations of 100 ns long unbiased simulations started from the second set of 16 newly projected structures. (\textbf{C}) Configurations from the initial ten unbiased simulations that were extended and are here tracked up to 2 $\mathrm{\mu s}$, (\textbf{D}) from 2 to 3 $\mathrm{\mu s}$ (black squares), and from from 3 to 4 $\mathrm{\mu s}$ (magenta squares). The free energy surface was previously extracted from a 2.52 ms long equilibrium simulation.}
\label{fig:globalDM}
\end{figure*}

\subsection*{Benchmark 2}

Having demonstrated the power of iMapD in applications to well-characterized model systems, we next use it to chart the configuration space of the biologically relevant Mga2 sensor of lipid saturation \cite{Covino2016}. For this challenging molecular system, even millisecond long atomistic MD simulations proved insufficient to observe a dimer dissociation event. However, they provide us with an excellent reference for the dimeric bound state \cite{Covino2016}. Fig. \ref{fig:globalDM} shows the corresponding free energy landscape as a function of the first two global DC of the dimers, with the 4 highly populated clusters corresponding to local minima. Importantly, we do not use this surface to guide iMapD in any way, only to give the reader a global view of the progress in the search.
%
%The Mga2 protein plays a key role in the regulation of lipid saturation levels in the yeast endoplasmic reticulum (ER). 
%
%Recent simulations and experiments identified a unique rotation-based  sensing mechanism to probe the membrane characteristics \cite{Covino2016}. In response to changes in lipid saturation, the 30-amino acids transmembrane helices (TMH) anchoring Mga2 into the ER were found to rotate relative to each other in an Mga2 dimer, driven in part by packing effects acting on bulky protruding  tryptophans. Just probing the rotational dynamics and charting the underlying free energy landscape required MD millisecond-long simulations feasible only with a coarse-grained (CG) description \cite{Marrink2007,Monticelli2008}. However, even on this long time scale, only the TMH contact could be sampled, with TMH dissociation expected to occur on time scales orders of magnitude longer. Therefore, even in more than 3-ms simulations of a simplified CG description, the relevant configuration space of the dimer could not be sampled exhaustively.

%Here we show that with our uncharted exploration (iMapD) not only the competing Mga2 bound states, but also the unbinding pathways can be discovered, simply by strategic initialization of otherwise fully unbiased MD trajectories. This reduces computational costs at least 1000-fold.

As the first step in iMapD, we run a burst of ten short (100-ns) unbiased simulations initiated from the starting configuration. The resulting trajectories sample its vicinity, but do not escape the local free energy minimum (black squares Fig.\ref{fig:globalDM} first panel). We use the structures along these simulations to detect the boundary in the local DC representation, and from there we project outward, building 16 new starting configurations.  From each of these we start another burst of ten 100-ns long unbiased simulations. Although most of these trajectories fall back to the starting cluster, many are able to escape from it, landing into new regions of the landscape, effectively discovering most of the highly populated clusters during the first phase of the expansion (blue circles Fig.\ref{fig:globalDM} panel A). In the spirit of building a growing map of the landscape, we combine all configurations sampled so far and repeat the boundary detection and projection in newly calculated local DC to obtain new starting configurations. In a second iMapD round starting from them, we already visit all relevant regions of the landscape (blue circle Fig.\ref{fig:globalDM} panel B).

For reference, we extend the initial unbiased simulations to estimate the time scales necessary to explore the landscape in a purely equilibrium approach. In Fig.\ref{fig:globalDM} panel C, we can see that after running ten simulations for $2\,\mathrm{\mu s}$ each, only one trajectory is able to leave the starting cluster. In order to discover all the remaining clusters, each simulation must be run for $4\,\mathrm{\mu s}$. 

We now represent the exploration process by using two angles $u$ and $v$ that describe the relative orientation of the two TMH in a Mga2 dimer (see inset lower panel Fig.\ref{fig:globalUV}). We again took advantage of already available long equilibrium simulations to calculate a reference free energy surface as a function of $u$ and $v$. Due to the identity of the two TMH, the surface, shown in Fig.\ref{fig:globalUV}, is approximately mirror symmetric with respect to the bisector.

In iMapD, the first short unbiased simulations sample structures where the two reference W10 face each other, and, consistently with what we saw in the global DC representation, are confined to the starting state. The first expansion leads to the discovery of two new states, which contain configurations where the W10 are far apart, pointing in one case to opposite directions and in the other in the same direction. Taking into account the symmetry of the surface, the last relevant state is discovered during the second expansion. Importantly, in a few configurations at this stage, the two TMH are actually separated, which represents a disassociation event of the dimer. This particular new configuration is then sampled for almost half of the time during a third expansion.

\section{Discussion}
In this work, we described, implemented and tested iMapD, a geometry-based, machine-learning inspired approach to accelerating the extraction of information from atomistic and stochastic simulators - in particular, the computation of effective FES. 

\begin{figure*}
\centering
\includegraphics[width=1.0\linewidth]{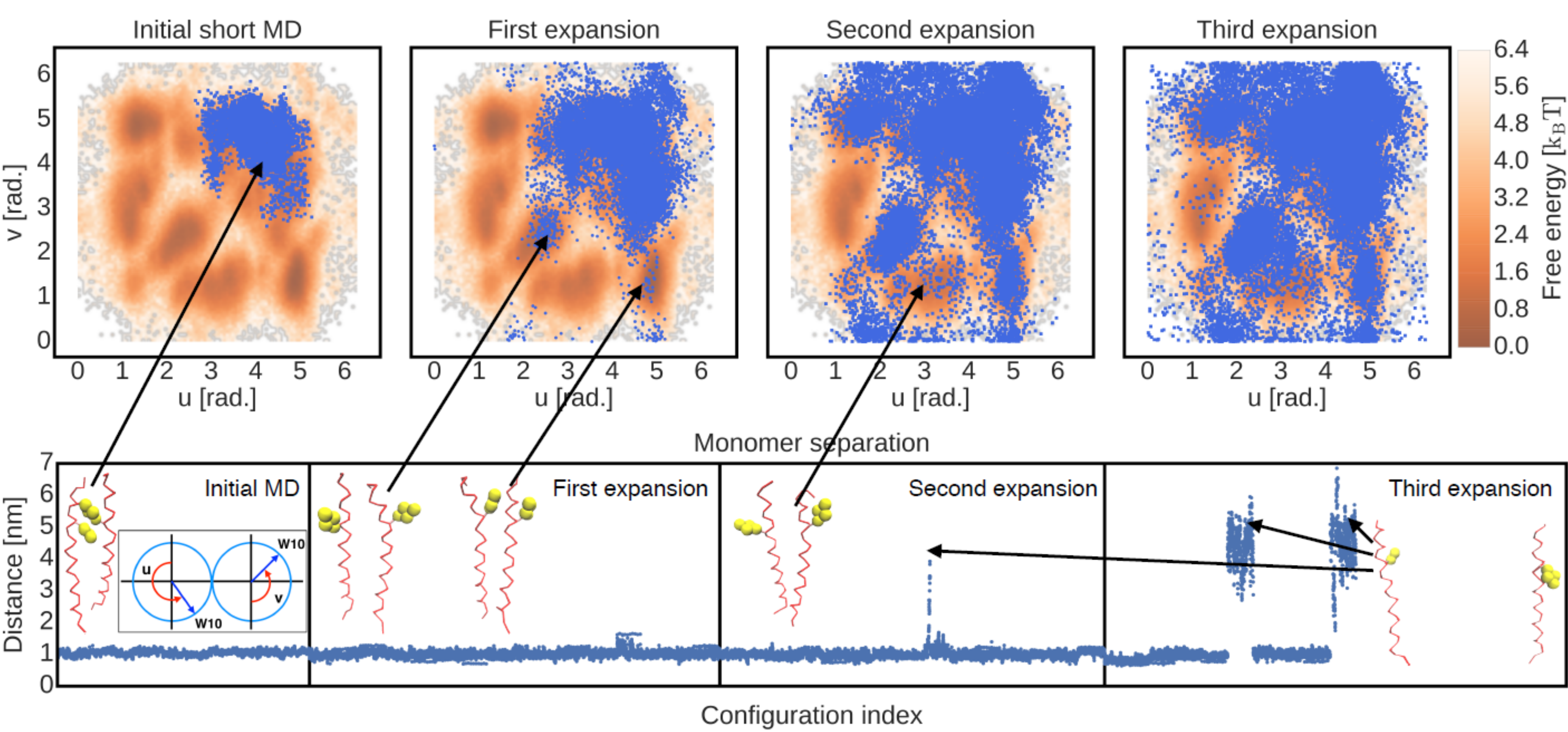}
\caption{(\textbf{Upper panel}) Enhanced exploration represented in $u, v$ space. Angles were calculated on configurations sampled  from cumulative trajectories simulated during the successive exploration phases and represented as blue squares. The free energy surface as a function of $u, v$ was extracted from a 2.52 ms long equilibrium simulation. (\textbf{Lower panel}) Distance separating the two TMH monitored during all the exploration phases. Insets show different representative structures of the dimer, with the reference residue W10 shown in yellow, and a schematic definition of $u$ and $v$.}
\label{fig:globalUV}
\end{figure*}

The algorithm has been tested on CG simulations of Mga2 TMH dimers, a system of biological relevance with rich conformational dynamics in the microsecond to millisecond regime and beyond; the two helices can make use of various contact interfaces, corresponding to the clusters in the free energy surfaces shown in Fig.\ref{fig:globalDM} and \ref{fig:globalUV}.

Our set-up mimics a situation in which only one structure is known, and MD simulations are restricted to short timescales due to the size of the system. On the one hand, the number of computing cores that can be used to parallelize a single simulation might be limited by lack of resources or bounds in the scaling behavior; on the other hand, the dynamics of complex %lipid bilayer
(bio)molecular systems are characterized by long correlation times. It stands to reason that in such situations of practical interest, running many short independent simulations is often more effective than focusing on few long ones. However, a crucial element of this strategy is to select appropriate initial configurations so not to get trapped in  configuration space. 

We have shown how by using machine learning algorithms (DMAPs and PCA) to infer new configurations from which to start bursts of short unbiased simulations we were able to efficiently discover new relevant structures of the Mga2 TMH dimers. Starting from a single initial structure, the iMapD algorithm was able, in only two iterations, to sample structures in the entire relevant configuration space of the dimer. We want to stress that all simulations that we have used are unbiased: after "intelligent" reinitialization no unphysical force was added to steer the dynamics of the system. 

In order to monitor the progress of the exploration, we used low dimensional free energy surfaces calculated as a function both of machine learning coordinates (DMAP), and physical variables (angles $u,\, v$). These surfaces are representations of the configuration space of the dimer, i.e., when both TMH are in close proximity ($\approx 1$ nm). As we saw in Fig.\ref{fig:globalUV} by monitoring the distance separating the two TMH, during the second and third expansion
(and thus after only a few tens of microseconds of cumulative simulated time), the algorithm sampled a dissociation event. 

In equilibrium simulations these occur on much longer time scales than the formation of the dimer itself. During more than 3 ms long equilibrium simulations of the Mga2 dimer, we never observed a single dissociation  event\cite{Covino2016}! 

This problem has been recently addressed in the context of the same CG model by using metadynamics \cite{Lelimousin2016}, where an unphysical history dependent bias must be added on the distance separating the two TMH, considered to be {\it a priori} a slow coordinate of the system. 

In our case, instead, the algorithm "discovers" this new extremely slow coordinate after having exhaustively explored the slow coordinates describing the conformational rearrangement in the dimer state shown in Fig.\ref{fig:globalDM} and \ref{fig:globalUV}. One can say that the algorithm gradually and adaptively discovers a hierarchy (an "atlas") of slow coordinates. 

Monitoring the actual dissociation events suggests a new slow variable: the relative tilt of the two helices.

The main attractive feature of the proposed approach is that it can explore low-dimensional effective free energy surfaces in high dimensional configurations spaces without the need of relying on {\em a priori} knowledge of suitable collective coordinates;
our coarse coordinates are progressively and adaptively revealed as computation progresses. 

The main assumption upon which our method is based is the same one that underpins most of the model reduction techniques in statistical mechanics: due to time scale separation, the system dynamics is mostly confined on low-dimensional (smooth) manifolds in phase-space \cite{ProcessesPaper,ChiavazzoPRE2011,EqFree1,EqFree2}.
Our approach squarely aims at exploiting smoothness of the low-dimensional manifolds which, for the gradient systems of interest here, act as the support of the free energy surface governing molecular and other atomistic dynamics.
Writing the expression {\em jumping as far as we trust the smoothness} above is, then, the pivot on which our approach "lives or dies". 
Two important issues, one relatively simple, and one deeper, determine the {\em tuning parameters} of our algorithm.
The first is the easier one: if the effective FES does not have hierarchical roughness, then there already exist two computational enabling technologies that support our algorithm. 
The first "enabler"  has to do with the local scaling of the noise through a {\em Mahalanobis like} distance \cite{Dsilva2015} which, combined with diffusion-map based data-mining conveniently factors out fast local oscillations ({\em curved} fast local invariant measures, "half-moons" as we call them in the discussion of \cite{halfmoons}.

The second "enabler" is straightforward: after factoring out these fast oscillations we have a smooth surface, and now we are faced with a numerical error control problem: the need for systematic adaptive step-size selection. 
We will not address this technical issue here; we simply note that the same computational machinery that, in traditional initial value problem solvers, allows one to make local error estimates can also be in principle used for our purposes. 
Performing the computation with one step, and then performing it twice with half the step, allows one to make a local "on line" error estimate and keep the computation below prescribed error bounds.

The second issue is deeper, and we will only pay lip service to resolving it, even though we believe that what we suggest is "the right way" to go about it. 
This is the critical (not for our relatively simple examples!) issue of hierarchical roughness. 
This implies (in SDE language) that our potentials are multi-scale potentials; and, possibly, that our noises may not be just additive.
Here, we revert to the discussion above about "what the best off-the-shelf estimation techniques" for multiscale diffusions, and maybe not only diffusions, but, say, Levy flight processes may be. 
In all our discussion, we assumed that the effective equation is a Langevin (or the associated Fokker-Planck). 
For simple "egg-carton" like potentials, as in the seminal work of Pavliotis and Stuart, it is possible through ingenious but relatively straightforward tools, like subsampling, to "go around"  the roughness, and estimate a smooth {\em effective} SDE \cite{pavliotis2007,Kalliadasis2015,Krumscheid2015,Calderon2007a}.
%%%YGK
%% just add one or two references from Calderon
%%

What is, however, more systematically missing (and missing, to significant extent, in the SDE estimation literature) is a round of data processing (and if necessary) additional data collection for {\em hypothesis testing}. 
In 2007, and in a more general context, we discussed this issue of "Deciding the Nature of the Coarse Equation through Microscopic simulations..." \cite{Li2007}. 
As the abstract of that papers states, "{\em ...The effective coupling of microscopic simulators with macroscopic behavior requires certain decisions about the nature of the unavailable coarse equation. ... In the absence of an explicit formula ..., we propose, implement, and validate a simple scheme for deciding these and other similar questions about the coarse equation using only the microscopic simulator.}".

This excerpt was chosen to suggest we have done a lot (and we believe there are contributions there for several types of problems); yet we believe that the collection of data for hypothesis testing about the nature of unavailable effective SDEs is a nascent field, and we are cognizant of relatively few efforts in this direction. 
Yet given the microscopic simulators, one can collect the data necessary for such algorithms, and we believe that, even though there will be technical difficulties, and good mathematics in the process, this is an area that will advance significantly in the near future, and our approach will benefit from these advances.

While discussing estimation, there is another significant (and, fortunately, less difficult) item to consider: the exploitation of the estimated local potential gradient -the heat map, or "color map", on our carpet- in {\em informing} our geometric exploration of the carpet. 
This has been, to a large extent, discussed in \cite{ThomasPaper} when the collective coordinates were known - and the issues remain the same, because they are, more or less, common-sense issues: when at the bottom of a well, we probably want to move upwards; we may want our steps to be along local geodesics; we may want our steps (the location of our "fresh clouds") to maybe try to conform to level sets of the effective free energy; if we find a saddle, we may "just" kick a little "on the other side", and let the simulation find the new well bottom by itself; if there are surfaces that look like the Grand Canyon (huge gradients in some directions) maybe we do not want to go above some level set because the simulation (and the molecule itself) would never get there in a person's lifetime. 
We appreciate that these and several other {\em common sense} decisions have to be implemented in an automated fashion in the code, if the code is to be useful; that is a matter of effort and resources. 
Plumed \cite{plumed01} (or Colvars \cite{Colvars}) goes quite a long way towards being a platform in which to incorporate what is done here, and what we envision being done in an automated fashion.
It is also important to recognize that one may have "exotic" carpets that change dimensionality as the exploration proceeds ($2$-dimensions narrowing to $1$-dimension and then maybe "opening back out" like a river delta, to $2$-dimensions - just like the shapes of some children's kites as shown in \cite{ChiavazzoPRE2011}- , or, say $3$-dimensions narrowing to $1$-dimension and then back again, like bar bells). We can in principle deal with that geometrically, but we will not discuss this further in this paper
beyond referring to the work of Belkin et al. and others \cite{Belkin2012}, \cite{Deutsch2016}.
The important issue of adaptively determining the dimensionality of low-dimensional surfaces in data mining has been discussed in \cite{Rohrdanz2011}.
%%
%As a remark of practical interest, the implementation of the algorithm is relatively simple, and can be carried out with a few hundred lines of Python or Mathlab code.

It is fitting to close the discussion by a quotation from the 2016 B. Peters review article
\cite{Peters}: "{\em However, the methods in this review share one overarching disadvantage. Human intuition remains the best source of trial coordinates and mechanistic hypotheses, and there is no procedure for having an epiphany. All current algorithms for optimizing reaction coordinates work
within the space of chosen trial coordinates.}"
Our work here attempts such a "computer assisted epiphany": by adaptively revealing the exploration geometry, and by exploiting its smoothness to guide further exploration;
it makes a step towards circumventing human intuition in the discovery phase. Yet the 
rationalization of what has been discovered in terms of physically interpretable
candidate coordinates is an important post-processing step and significantly augments
the overall value of the process (e.g. confirming that, in the last stages,
the relative tilt between the two helices is "one-to-one" with the machine-discovered
coordinates). A useful discussion of the "man-versus-machine" detected variables
can be found in \cite{Frewen2011,Sonday2009}.
%%%

\section{Materials and methods}
\subsection*{DMAP: Mapping from ambient space to reduced space}
The data mining step of our procedure has been performed by the Diffusion Map (DMAP) method.
Compared to the popular Principal Component Analysis (PCA) \cite{Jolliffe}, DMAP enables us to extract a nonlinear embedding thus minimizing the dimension of the low-dimensional description.  
Full details on DMAPs can be found in \cite{Coifmanpaper1,LafonTh}. 
%%%YGK
%%  please add the Thesis of Lafon
%%
%
Let $\left(Y_1,...,Y_m\right)$ be a set of $m$ points in a $p$-dimensional (ambient) space.
To fix ideas, let $Y_i$ be an array collecting the $x,y,z$ Cartesian coordinates of all atoms (or beads in coarse-grained calculations) forming the $i-th$ configuration extracted from MD simulators. 
Let a {\em dissimilarity function} $d_{ij}=d_{ji}$ be defined between any pair of points $Y_i$ and $Y_j$, such that $d_{ij}=0$ only when points coincides, while the more dissimilar the point the larger $d_{ij}$.
An obvious choice (although not the only one) for $d_{i,j}$ is the Euclidean distance.
When dealing with molecular configurations, distances are to be evaluated upon removal of rigid translation, rotation and all other possible symmetries.
Based on $d_{ij}$, a pair-wise affinity function can be constructed, $w_{ij}=w\left(d_{ij}\right)$, where $w$ is monotonically decreasing and non-negative with $d>0$, while $w(0)=1$.
As usual in the DMAP literature, we utilized the heat kernel,
%
%\begin{equation}
$	w_{ij}=exp\left[ -\left( \frac{d_{ij}}{\epsilon} \right)^2 \right]$,
%\end{equation}
%   
where the model parameter $\epsilon$ is introduced in order to discriminate between points that are effectively linked from points that are not (i.e. those that are separated by a distance larger than $\epsilon$).
Moreover, we made use of the density invariant normalization of the matrix $W=\{ w_{ij} \}$. 
More details on the choice of the eigenvectors can be also found in \cite{ProcessesPaper}.

\subsection{Lifting from reduced to ambient space by LPCA}
Let $B$ be an arbitrary boundary point in the $p$-dimensional ambient space, which we want to extend outwardly with respect to the previously simulated (available) point cloud. 
Let us identify the $(n-1)$ nearest neighbors of $B$ in the ambient space.
Let $X$ be the $n \times p$ data matrix collecting the Cartesian coordinates of $n$ points, namely the ones within the chosen neighborhood of $B$, including $B$ itself.

Let us now perform Principal Component Analysis - PCA - of the matrix $X$. 
This yields:
\begin{itemize}
	\item The $p \times p$ matrix of loading: $C$;
	\item The $n \times p$ matrix of principal component scores: $S$;
	\item A vector $l$ with the $p$ eigenvalues of the covariance matrix $X$.
\end{itemize}
An estimate of the local dimension - $d_{loc}$- of data (i.e. in the chosen neighborhood of $\mathbf{B}$) can be readily obtained by setting a threshold for the maximum variance to keep as follows:
\begin{equation}\label{variance}
	\sum_{i=1}^{d_{loc}}l(i)/\sum_{j=1}^{p}l(j) > threshold.
\end{equation}
%%% GH: correct
with $l(i+1)\le l(i)$.
Alternatively, the dimension of the reduced space ($d_{loc}$) can be also fixed {\em a priori} once and forever at any boundary point. 
Regardless of the method we decide to use, let us assume that the above dimension $d_{loc}$ is known.
Let us consider the $n \times d_{loc}$ matrix $Y$ collecting the reduced PCA coordinates of the $n$ points of interest. 
In other words, $Y$ is provided by the first $d_{loc}$ columns of the above matrix $S$. 
Let $y_B$ and $y_{center}$ be the PCA reduced coordinates of the above boundary point $B$ and of the center of mass of the considered neighborhood, respectively.
We compute the following (row) unit vector $v$ in PCA reduced space along which we intend to project outward the boundary point $B$.
\begin{equation}
	v=\frac{{y_B-y_{center}}}{{\left| {y_B - y_{center}} \right|}}
\end{equation}
A new point in reduced PCA space can be identified as
%
%\begin{equation}
$	y_{new} = y_B + c v$,
%\end{equation}
%
where $c$ is a non-negative scalar quantity stipulating how far we intend to extend the point $B$ from the current location.
Lifting of the new point $y_{new}$ into ambient space can be readily accomplished by a linear mapping, 
%
%\begin{equation}
$Y_{new} = y_{new} \tilde{C} + \bar{X} $
%\end{equation}
%
where the $d_{loc} \times p$ matrix $\tilde{C}$ is given by the (transposed) first $d_{loc}$ columns of the above matrix $C$ (i.e. the matrix of loadings), while $X$ is the mean row vector, where each of the $p$ (ambient space) coordinate is averaged over the $n$ points in the chosen neighborhood.

\subsection*{Computations with Ala dipeptide}
The Ala dipeptide is simulated with GROMACS 4.5.5 \cite{GROMACS1,GROMACS2} in a periodically replicated box with dimensions of $2 \times 2 \times 2$ $nm^3$. Solvent is treated implicitly, using the Still generalized Born formalism with a cut-off of 0.8 nm.
The temperature is maintained constant at $300 K$ by means of velocity rescaling thermostat \cite{vrescale}.

When searching for DMAP low-dimensional embeddings, all configurations are first aligned to a reference configuration using the Kabsch algorithm \cite{Kabsch76,Kabsch78}, and afterward the standard Euclidean distance is used as pair-wise dissimilarity function.
The DMAP model parameter was set at $\epsilon=0.35$ nm.
When performing local PCA, at each boundary point $n=65$ nearest neighbors are considered, whereas the local dimension $d_{loc}$ is automatically estimated by setting a threshold for maximum variance of $0.95$.

The non-negative scalar quantity $c$ for local extension is chosen in the range: $0.05<c<0.12$.
Starting from each new extended configuration, 2 short bursts are performed by each time randomly re-assigning  Maxwell-Boltzmann velocities to atoms.
Simulation bursts consist of 15000 simulation steps with a time step of $0.02$ fs.
The latter unusually small time step is not essential for computations: it was chosen for convenience as it ensured a sufficiently large number of samples along the burst trajectories.

\subsection*{Computations with Mga2}
\subsection*{Model and simulation details}
The 30 amino-acids long transmembrane domain of Mga2 (sequence in single letter code: RNDKMLIFFWIPLTLLLLTWFIMYKFGNQD ) was modeled as an alpha helix in the MARTINI v2.2 force field \cite{Marrink2007,Monticelli2008}. We used the insane tool \cite{Wassenaar2015} to assemble for each simulation a 10x10x10 nm box containing two Mga2 monomers, about 300 POPC lipids, water and ion beads corresponding to a 0.15 M NaCl concentration, for a total amount of about 10,000 beads.

Each initial configuration was relaxed by using 15000 steps of steepest descent, and then equilibrated for 2 ns at a temperature of 303 K and pressure of 1 atm, restraining the positions of the protein beads rescaled compatibly with the pressure coupling. Temperature was kept constant with the velocity rescaling thermostat \cite{vrescale} and pressure with the semiisotropic Berendsen barostat \cite{Berendsen1984} during equilibration, and the semiisotropic Parrinello-Rahman barostat \cite{Parrinello1980} during the production runs.

All simulations were performed in GROMACS 4.6.7 \cite{Abraham2015,GROMACS1,GROMACS2,Pronk2013}, using a time step of 20 fs.

\subsection*{Enhanced sampling details}
We initially ran 10 independent simulations starting from the same structure, each 100 ns long. We saved configurations containing only the protein degrees of freedom every 2 ns, and aligned them in a self-consistent way to the average sampled configuration, removing translations and rotations with the Kabsch algorithm \cite{Kabsch76,Kabsch78}. In particular, we first aligned the trajectory to an arbitrary configuration, calculated the average configuration and used it to align the trajectory again, repeating the procedure until the RMSD between two consecutive average configurations was smaller than 0.01 nm. The alignment was done on the backbone beads of residues 3-28 of each monomer. Furthermore, we took into account the identity of the two monomers, which introduces an exchange-symmetry in the system. We thus considered for every frame the structure with the smallest RMSD to the reference upon swapping of the two monomers.

We calculated the first two DC in the Cartesian space of the aligned configurations, using the Euclidean metric and $\epsilon=5$ nm, and approximated the boundary of the obtained points with a convex hull. We projected each point on the boundary outwards at a distance of $v=5$ nm along the local PCA, which were calculated on its 100 nearest neighbors, with $d_{loc}$ chosen to keep $95\%$ of the original variance. 

We added lipids around the projected dimer and solvated the resulting bilayer, then shortly equilibrated the system, obtaining 16 new configurations (Table \ref{tab:table}). We then ran 10 independent 100 ns long unbiased simulations for each new configuration by randomly initializing the initial velocities, and merged the new trajectories to the initial ones.

After this first expansion, we repeated the entire procedure for a second and a third time (expansions 2 and 3) obtaining, respectively, 16 and 12 new configurations (Table \ref{tab:table}). Newly discovered configurations where the two monomers are separated were excluded from the procedure, since they would dominate the representation in DC.

\subsection*{Global DM calculation}
The free energy surface as a function of the first two global diffusion coordinates shown in Fig.\ref{fig:globalDM} was calculated using 24,815 configurations of the dimer sampled at equal times from a previously reported 2.52 ms long equilibrium trajectory \cite{Covino2016} that was self-consistently aligned to the average configuration as already explained.

To represent newly sampled configurations on the global DM landscape, we aligned them on the same average configuration described above and combined them to the configurations sampled from the long equilibrium trajectory, hence evaluating the DM on the combined set for each new trajectory. All DM were calculated using the Euclidean metric and $\epsilon=5$ nm.

\subsection*{Relative orientation calculation}
The two angles $u$ and $v$ used in Fig.\ref{fig:globalUV} define the relative rotation of the two alpha helices forming a Mga2 dimer. $u$ is the angle defined counter-clockwise between the orthogonal line to the direction connecting the centers of mass of the two helices and the vector pointing to residue W10 of the first monomer; $v$ is the angle defined counter-clockwise between the orthogonal line to the direction connecting the centers of mass and the vector pointing to residue W10 of the second monomer (see inset in Fig.\ref{fig:globalUV}). Both angles have periodicity $2\pi$. For chemically and structurally identical monomers, we would have mirror symmetry with respect to the line $u-v=2\pi$. 

The reference free energy landscape of Fig.\ref{fig:globalUV} was calculated by using 248,144 frames sampled at equal times from a 2.52 ms long equilibrium trajectory \cite{Covino2016}. We calculated $u$ and $v$ both in the reference of the first monomer, and in the reference of the second monomer, in this way effectively enforcing the monomer-exchange symmetry of the system. Any residual deviation from the mirror symmetry in the surface is due to some flexibility of the helices, which can be only approximately considered as rigid bodies.

Analysis and visualization of the data were performed with NumPy \cite{VanderWalt2011}, SciPy \cite{Jones}, IPython \cite{Perez2007}, Matplotlib \cite{Hunter2007} and MDAnalysis \cite{Michaud-Agrawal2011}. Molecular representations were made with VMD \cite{Humphrey1996, Stone1995}.

\section*{Appendix}
\begin{table*}%[tbhp]
\centering
\caption{Summary of performed MD simulations of the Mga2 dimer}
\begin{tabular}{lccccc}
 &  \begin{tabular}[c]{@{}c@{}}Nr. of starting\\structures\end{tabular}   & Nr. of simulations & \begin{tabular}[c]{@{}c@{}}Cumulative\\simulation time [$\mu s$]\end{tabular} & $u, v$ discovered states & \begin{tabular}[c]{@{}c@{}}Cumulative count of \\ dissociation events\end{tabular} \\
%\midrule
1. Initial unbiased & 1 & 10 & 1 & 1/4 & 0\\
2. Expansion 1 & 16 & 160 & 16 & 3/4 & 0\\
3. Expansion 2 & 16 & 150 & 15 & 4/4 & 1\\
4. Expansion 3 & 12 & 100 & 12 & 4/4 & 1\\
5. Reference unbiased & 1 & 10 & 40 & 4/4 & 0\\
6. Equilibrium free energy & 10 & 10 & 2,520 & 4/4 & 0\\
%\bottomrule
\end{tabular}
\label{tab:table}
\end{table*}
\begin{figure}
\centering
\includegraphics[width=.8\linewidth]{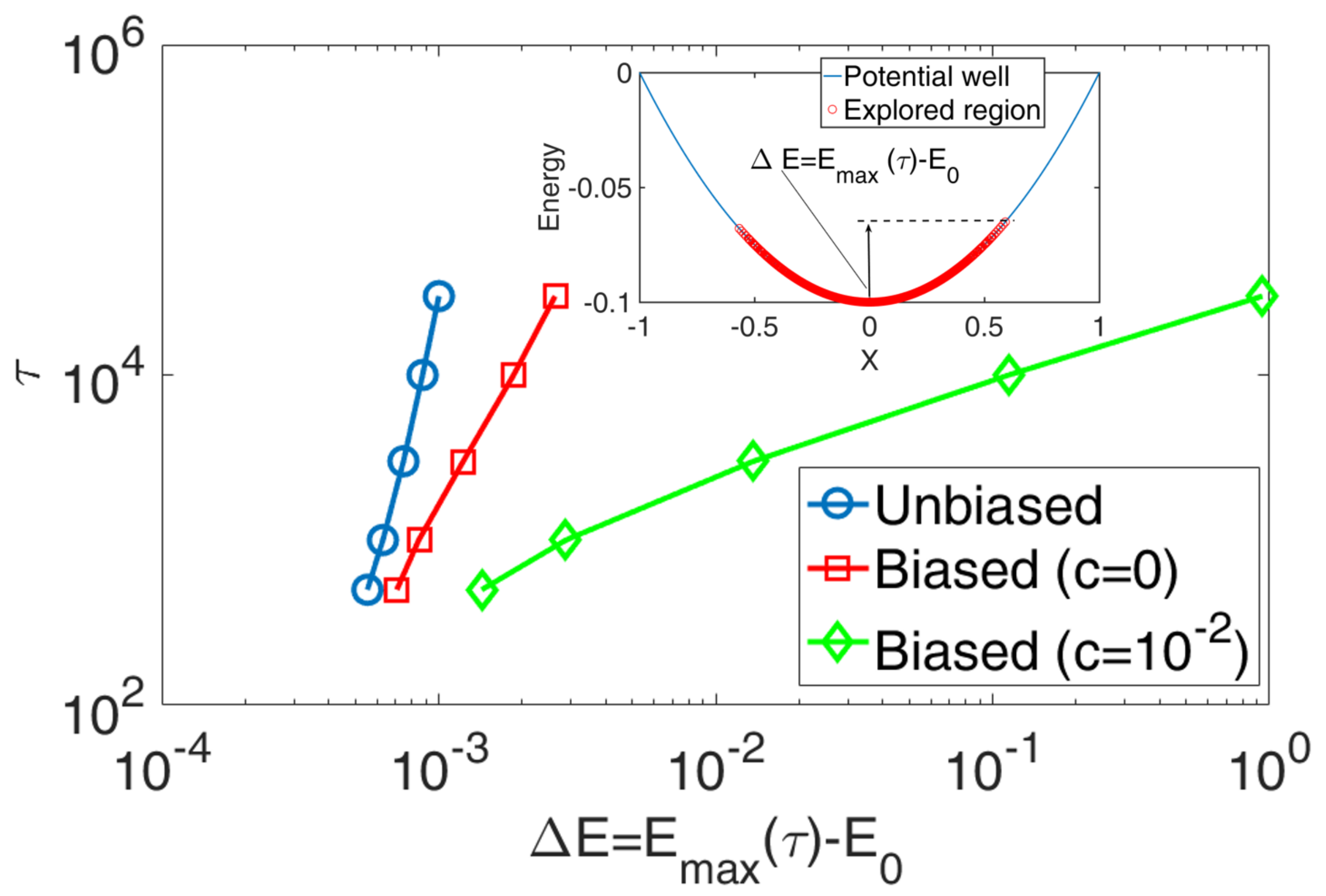}
\caption{\textbf{Inset.} An SDE with harmonic potential whose minimum is at $x=0, E_0=-0.1$: After a simulation time $\tau$ the explored region has the (expected) maximum height in energy $E_{max}(\tau)$ shown. The figure compares results of unbiased simulations with those biased by (a) simple reinitialization at the edges and (b) geometrically biased reinitializations, through the outward extensions advocated here. The unbiased SDE \ref{sdeeq} has been solved using a Euler-Maruyama scheme with $dt=0.5$. We caution the reader that this is just an illustrative caricature.}\label{SDEcomp}
\end{figure}
Let us consider the stochastic dynamics described by 
\begin{equation}%\label(1DSDE)
	dx = \left(- \frac{\partial V}{\partial x} dt + D\sqrt{2} dW \right)
\end{equation}\label{sdeeq}
with harmonic potential $V\left(x\right) = \frac{E_0}{\sigma_0^2} x^2-E_0$.
for $E_0=0.1$, $D=0.01$ and $\sigma_0=1$.
For unbiased simulations initialized around the bottom of the energy well, the expected time to 
visit a given region at least once grows exponentially with the spanned energy difference, as shown by open circles in the inset of Figure \ref{SDEcomp}).  
For computations biased by re-initializing at the edges, $x_1=min(x)$ and $x_2=max(x)$, of the interval fathomed after a simulation time $\tau=100$, in the spirit of the approach in \cite{CeciliaPaper} a significant advantage can be already noticed as shown in  Figure \ref{SDEcomp} by the open squares.
However, remarkable additional computational speedup (i.e. several orders of magnitude) can be achieved if the solver of \ref{sdeeq} can be reinitialized each time {\em extending beyond} the already  explored domain.
For an extension step of only $c=10^{-2}$ (re-initializing each time at $x_1= min(x) - c, 
x_2= max(x) + c$ and then running unbiased for $\Delta t=100/2$ at each end, the improvement is clearly visible in the open diamonds in the Figure.
%%%

\section{Author contributions}
EC and IGK initially planned the work. ASG, CWG and RRC contributed to the data mining aspects of the problem. EC (benchmark 1) and RC with GH (benchmark 2) performed the computations shown. EC and IGK with assistance from all authors wrote the paper.

\section*{Acknowledgments}
%An Acknowledgments section, if used, immediately precedes the References. Sponsorship and financial support acknowledgments should be included here.
This work was partially supported by the US National Science Foundation, the US AFOSR (Dr. Darema) and DARPA (IGK). E.C. acknowledges partial support of Italian Ministry of Education through the NANO-BRIDGE project (PRIN 2012, grant number 2012LHPSJC). R.C. and G.H. were supported by the Max Planck Society. E.C., R.C. and I.G.K. also wish to acknowledge the hospitality and support of IAS-TUM in Garching.

\section*{References}
%
%\bibliography{REFER}
\end{document}